\documentclass{article}

\usepackage[utf8]{inputenc}
\usepackage[T1]{fontenc}
\usepackage{lmodern}
\usepackage{microtype}

\usepackage{amsmath, amssymb, amsthm}

\usepackage[margin=2.5cm,headheight=20pt]{geometry}
\usepackage{fancyhdr}
\usepackage{graphicx}
\usepackage{subcaption}
\usepackage{float}
\usepackage{framed}

\usepackage{xcolor}
\definecolor{sectioncolor}{HTML}{2c3e50}         
\definecolor{linkcolor}{HTML}{1565c0}            
\definecolor{keywordcolor}{rgb}{0.7, 0.1, 0.1}   
\definecolor{tacticcolor}{rgb}{0.7, 0.1, 0.1}    
\definecolor{commentcolor}{rgb}{0.4, 0.4, 0.4}   
\definecolor{symbolcolor}{rgb}{0.0, 0.1, 0.6}    
\definecolor{sortcolor}{rgb}{0.1, 0.5, 0.1}      
\definecolor{attributecolor}{rgb}{0.7, 0.1, 0.1} 
\definecolor{declnamecolor}{HTML}{6f42c1}        
\definecolor{abstractbg}{HTML}{f8f9fa}           

\usepackage{titlesec}
\titleformat{\section}
  {\fontsize{14}{17}\bfseries\color{sectioncolor}}
  {\thesection}{1em}{}
\titleformat{\subsection}
  {\fontsize{12}{14}\bfseries\color{sectioncolor}}
  {\thesubsection}{1em}{}
\titleformat{\paragraph}[runin]
  {\bfseries\color{sectioncolor}}
  {}{}{}[.]
\titlespacing*{\section}{0pt}{1.5\baselineskip}{0.5\baselineskip}
\titlespacing*{\subsection}{0pt}{1.2\baselineskip}{0.4\baselineskip}
\titlespacing*{\paragraph}{0pt}{0.8\baselineskip}{0.5em}

\usepackage{listings}

\usepackage[numbers,sort&compress]{natbib}

\usepackage{xspace}

\usepackage[breaklinks=true,colorlinks=true,linkcolor=sectioncolor,citecolor=linkcolor,urlcolor=linkcolor]{hyperref}
\usepackage{xurl} 

\usepackage{cleveref}

\newcommand\themaintitle{LeanArchitect}
\newcommand\thesubtitle{Automating Blueprint Generation for Humans and AI}

\makeatletter
\renewcommand{\maketitle}{%
  \begin{center}
    \vspace*{1em}
    {\fontsize{28}{34}\selectfont\bfseries\themaintitle\par}
    \vspace{0.6em}
    {\fontsize{14}{18}\selectfont\thesubtitle\par}
    \vspace{1.5em}
    {\large\begin{tabular}[t]{c}
      Thomas Zhu\textsuperscript{1} \quad Pietro Monticone\textsuperscript{2} \quad Jeremy Avigad\textsuperscript{1} \quad Sean Welleck\textsuperscript{1}
    \end{tabular}\par}
    \vspace{0.8em}
    {\small\textsuperscript{1}Carnegie Mellon University \quad \textsuperscript{2}University of Trento\par}
    \vspace{1.5em}
  \end{center}
}
\makeatother

\renewcommand{\headrulewidth}{0.4pt}
\renewcommand{\headrule}{\hbox to\headwidth{\color{sectioncolor}\leaders\hrule height \headrulewidth\hfill}}

\newcommand{\leanarchitect}{LeanArchitect\xspace}
\newcommand{\leanblueprint}{leanblueprint\xspace}

\definecolor{codebg}{HTML}{f6f8fa}

\lstdefinestyle{leanstyle}{
  showstringspaces=false,
  keepspaces=true,
  morestring=[b]",
  morestring=[d],
  tabsize=3,
  extendedchars=false,
  sensitive=true,
  breaklines=true,
  breakatwhitespace=true,
  basicstyle=\ttfamily\small,
  captionpos=b,
  columns=[l]fullflexible,
  identifierstyle={\ttfamily\color{black}},
  keywordstyle=[1]{\ttfamily\color{keywordcolor}},
  keywordstyle=[2]{\ttfamily\color{sortcolor}},
  keywordstyle=[3]{\ttfamily\color{tacticcolor}},
  keywordstyle=[4]{\ttfamily\color{attributecolor}},
  stringstyle={\ttfamily\color{symbolcolor}},
  commentstyle={\ttfamily\footnotesize\color{commentcolor}},
  escapechar=§,
  backgroundcolor=\color{codebg},
  frame=single,
  framerule=0pt,
  xleftmargin=0.5em,
  xrightmargin=0.5em,
  framexleftmargin=0.3em,
  framexrightmargin=0.3em,
}
\newcommand{\declname}[1]{{\ttfamily\color{declnamecolor}#1}}

\definecolor{quotebar}{HTML}{3498db}
\definecolor{quotebg}{HTML}{f0f7fc}
\usepackage{mdframed}

\mdfdefinestyle{abstractstyle}{
  backgroundcolor=abstractbg,
  linewidth=0pt,
  innerleftmargin=12pt,
  innerrightmargin=12pt,
  innertopmargin=10pt,
  innerbottommargin=10pt,
  skipabove=0.5em,
  skipbelow=1em,
}
\renewenvironment{abstract}{%
  \begin{mdframed}[style=abstractstyle]
  {\small\bfseries\color{sectioncolor}Abstract}\par\vspace{0.4em}
  \small
}{%
  \end{mdframed}
}

\mdfdefinestyle{quotestyle}{
  linewidth=3pt,
  linecolor=quotebar,
  backgroundcolor=quotebg,
  topline=false,
  bottomline=false,
  rightline=false,
  leftline=true,
  innerleftmargin=12pt,
  innerrightmargin=12pt,
  innertopmargin=10pt,
  innerbottommargin=10pt,
  skipabove=\baselineskip,
  skipbelow=\baselineskip,
}
\renewenvironment{quote}{\begin{mdframed}[style=quotestyle]\itshape}{\end{mdframed}}

\newcommand\latexcode{\lstinline[language={[latex]tex},keywordstyle=\ttfamily\color{black}]}
\newcommand\bplatexcode{\lstinline[language={[latex]tex},keywordstyle=\ttfamily\color{keywordcolor},moretexcs={inputleannode,inputleanmodule}]}
\newcommand\leancode{\lstinline[language=lean]}
\newcommand\bashcode{\lstinline[language=bash, %
  literate=%
  {/}{}{0\discretionary{/}{}{/}}%
  {\_}{}{0\discretionary{\_}{}{\_}}%
  {.}{}{0\discretionary{.}{}{.}}]}

\begin{document}

\maketitle

\begin{abstract}%
\noindent Large-scale formalization projects in Lean rely on blueprints: structured dependency graphs linking informal mathematical exposition to formal declarations. While blueprints are central to human collaboration, existing tooling treats the informal (\LaTeX{}) and formal (Lean) components as largely decoupled artifacts, leading to maintenance overhead and limiting integration with AI automation. We present \leanarchitect, a Lean package for extracting, managing, and exporting blueprint data directly from Lean code. \leanarchitect introduces a declarative annotation mechanism that associates formal declarations with blueprint metadata, automatically infers dependency information, and generates \LaTeX{} blueprint content synchronized with the Lean development. This design eliminates duplication between formal and informal representations and eases fine-grained progress tracking for both human contributors and AI-based theorem provers. We demonstrate the practicality of \leanarchitect through the automated conversion of several large existing blueprint-driven projects, and through a human--AI collaboration case study formalizing a multivariate Taylor theorem. Our results show that \leanarchitect improves maintainability, exposes latent inconsistencies in existing blueprints, and provides an effective interface for integrating AI tools into real-world formalization workflows.
\end{abstract}

\newpage

\section{Introduction}

A long-term goal of formal mathematics is to represent mathematical knowledge in a precise, machine-checked language that supports both verification and large-scale collaboration. Proof assistants such as Lean~\cite{demoura2015lean, demoura2021lean} have made substantial progress toward this goal, with libraries like Mathlib~\cite{The_mathlib_Community_2020} formalizing much of undergraduate mathematics and an increasing body of research-level material.

As formalization projects scale, managing structure becomes as important as managing proofs. Large developments are typically decomposed into many interdependent definitions and theorems, spread across multiple files and contributors. To support this process, many Lean projects adopt \emph{blueprints}~\cite{massot_leanblueprint}: high-level dependency graphs that describe what remains to be formalized and how results depend on one another. Blueprints serve both as planning documents and as coordination tools for distributed teams.

Existing blueprint workflows, however, exhibit two fundamental limitations. First, blueprint information is typically duplicated across an informal \LaTeX{} document and the formal Lean code. As proofs evolve, this duplication creates significant maintenance overhead and opportunities for divergence. Second, current tooling provides limited support for integrating AI tools, despite the fact that blueprints naturally decompose formalization tasks into small, well-scoped units suitable for automation. These limitations have become more pressing as AI-based theorem provers mature. While modern systems can solve many isolated problems, real-world formalization requires reasoning about dependencies, partial progress, and natural language proofs---precisely the information encoded in blueprints. Without a native connection between blueprints and Lean code, deploying such systems in large projects remains limited in scope.

This paper introduces \emph{\leanarchitect}, a Lean-native framework for blueprint extraction and management. \leanarchitect allows developers to annotate Lean declarations with blueprint metadata, automatically infers dependency and proof-status information, and generates synchronized \LaTeX{} blueprint content directly from the Lean environment. By unifying the formal and informal views of a project, \leanarchitect reduces duplication, improves consistency, and provides a structured interface for AI-assisted formalization.

The contributions of this paper are:
\begin{itemize}
\item An annotation and extraction mechanism for blueprint nodes in Lean,
\item Automatic inference of dependencies and proof status in Lean,
\item Integration with existing \LaTeX{} blueprint workflows,
\item Empirical validation through conversion of large existing projects and a human--AI formalization case study.
\end{itemize}

\leanarchitect does not replace the existing \leanblueprint tool~\cite{massot_leanblueprint}; rather, it complements \leanblueprint by synchronizing Lean data automatically to the blueprint.

\leanarchitect is available as a Lean 4 library at \url{https://github.com/hanwenzhu/LeanArchitect}.

\section{Related Work}

\paragraph{Blueprints in Lean}
The term blueprint originates from Patrick Massot's \leanblueprint project~\cite{massot_leanblueprint}, a plas\TeX{}-based system that generates dependency graphs from \LaTeX{} documents using macros such as \latexcode{\uses} and \latexcode{\leanok}. Leanblueprint has been widely adopted in non-Mathlib projects due to its effectiveness in organizing large formalization efforts (such as~\cite{fvp_carleson, buzzard_flt, riehl_infinity-cosmos, kontorovich_pnt, degenne2025brownian, tao2025equational}). Although \leanblueprint already has a \leancode{checkdecls} command for checking whether all formal declarations in the blueprint have a corresponding Lean part, there is no support for syncing dependency and proof status information. \leanarchitect builds on this ecosystem by providing Lean-native extraction of blueprint data, reducing manual synchronization between \LaTeX{} and Lean.

\leanarchitect is also inspired by Ian Jauslin and Alex Kontorovich's \leanblueprint-extract tool,\footnote{See the \href{https://github.com/AlexKontorovich/PrimeNumberTheoremAnd/tree/d89e0d9/leanblueprint-extract}{GitHub repository}.} which is an extension of \leanblueprint that extracts raw comments from Lean to generate the \LaTeX{} blueprint. However, \leanblueprint-extract is limited in that the user has to manually write metadata such as \latexcode{\uses} dependencies, and the blueprint structure is restricted to follow the Lean code structure exactly.

\paragraph{Data extraction tools for Lean}
The tool doc-gen4\footnote{See the \href{https://github.com/leanprover/doc-gen4}{GitHub repository}.} generates HTML documentation directly from Lean code. While its purpose is different, \leanarchitect's code is inspired by doc-gen4. The tool ntp-toolkit~\cite{hu2024minictx} extracts declarations and unfinished proofs from Lean, primarily for training and testing AI tools like LeanHammer~\cite{zhu2025premise}. However, these systems focus on documentation or analysis rather than blueprint-driven project management, and they do not capture informal mathematical exposition or dependency structure at the level required for blueprint workflows.

\paragraph{AI-assisted theorem proving}
Neural theorem provers have advanced rapidly, with capability increasing from simple high-school problems to complex undergraduate problems~\cite{zheng2021minif2f, yang2023leandojo, xin2024deepseek, xin2024deepseek1.5, ren2025deepseek, lin2025goedel, lin2025goedel2, wang2025kimina, chen2025seed, chen2025seed1.5, achim2025aristotle, hubert2025olympiad}. They are increasingly tested on and integrated into real-world problems such as miniCTX~\cite{hu2024minictx} and formal-conjectures~\cite{deepmind_formalconjectures}. However, most setups consist largely of isolated problems and do not reflect the dependency-rich blueprint structure of real formalization projects. Most recently, RLMEval~\cite{poiroux2025rlmeval} incorporates blueprint-style information for evaluation. However, most prior work does not address the tooling required to deploy AI systems within existing blueprint-based Lean developments. \leanarchitect fills this gap by exposing blueprint structure directly in Lean, enabling AI systems to reason about partial progress, dependencies, and informal context within large projects.

\section{Methods}

\subsection{Overview}

\leanarchitect is a tool for extracting blueprint information directly from Lean code. Its core design principle is to minimize duplication between \LaTeX{} and Lean, by treating Lean as the authoritative source of information for any formalized blueprint node.

The system introduces a new attribute, \leancode{@[blueprint]}, which can be attached to Lean definitions and theorems. This attribute records metadata such as a \LaTeX{} label, natural language statements and proofs, and project-management annotations. In addition, \leanarchitect automatically infers metadata such as:
\begin{samepage}
\begin{itemize}
\item Dependencies between definitions and theorems,
\item Formalization status of the theorems (i.e., if they are \leancode{sorry}-free).
\end{itemize}
\end{samepage}

\begin{figure}[ht]
  \centering
  \begin{subfigure}{\textwidth}
  \includegraphics[page=1,width=\textwidth]{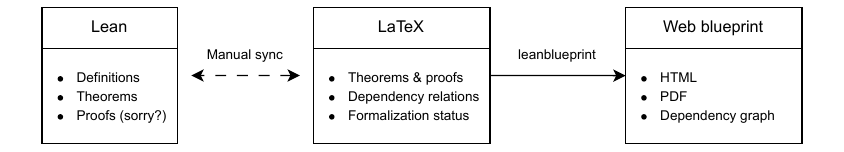}
  \caption{Blueprint generation workflow without \leanarchitect.}
  \end{subfigure}

  \bigskip

  \begin{subfigure}{\textwidth}
  \includegraphics[page=2,width=\textwidth]{figs/LeanArchitect.drawio.pdf}
  \caption{Blueprint generation workflow with \leanarchitect.}
  \end{subfigure}

  \caption{Comparison of blueprint generation workflows with and without using \leanarchitect. (a) Without \leanarchitect, the entire \LaTeX{} blueprint needs to be manually written and synchronized with the evolving formalization part. (b) With \leanarchitect, maintainers only need to manually write the structure of the \LaTeX{} blueprint, whose dependency relations and formalization status are automatically synchronized from the corresponding Lean part.}
  \label{fig:leanarchitect-workflow}
\end{figure}

Blueprint data is stored in an environment extension and can be exported via a dedicated Lake facet. The exported artifacts consist of \LaTeX{} fragments, which integrate seamlessly with existing \leanblueprint documents through macros such as \bplatexcode{\inputleannode}. See \Cref{fig:leanarchitect-workflow} for a diagram.

This design supports a workflow in which informal exposition and formal proofs evolve together, while minimizing manual duplication and enabling structured interaction with AI automation.

\subsection{Blueprint Attribute}

\leanarchitect provides a new Lean attribute \leancode{@[blueprint]} that can be attached to definitions and theorems. Users can optionally supply metadata such as a \LaTeX{} label, natural language statement and proof, and other project-management annotations. The attribute serves as the user interface between Lean code and the blueprint system. A typical example is:

\begin{lstlisting}
@[blueprint "thm:add-comm"
  (statement := /-- Addition in $ℕ$ is commutative. -/)]
theorem §\declname{MyNat.add\_comm}§ (a b : MyNat) : a + b = b + a := by
  /-- By induction and then \cref{lem:zero-add, lem:succ-add}. -/
  induction a with
  | zero => exact b.zero_add
  | succ a ih => sorry_using [MyNat.succ_add]
\end{lstlisting}

\subsection{Environment Extension}

After tagging a declaration, \leanarchitect constructs an internal representation of the corresponding blueprint node and stores it in an environment extension \leancode{blueprintExt}. For each tagged declaration, \leanarchitect automatically infers dependency information by recursively traversing the constants used in the type and value of the declaration, and determines proof status by checking if \leancode{sorry} is used. Then an internal \leancode{Node} is constructed; e.g.~with the following data:

\begin{lstlisting}
-- Identifiers
name := MyNat.add_comm, latexLabel := "thm:add-comm"
-- Statement status, dependencies, and text
statement := { leanOk := true, uses := ["def:nat"],
              text := "Addition in $ℕ$ is commutative." }
-- Proof status, dependencies, and text
proof := { leanOk := false, uses := ["lem:zero-add", "lem:succ-add"],
         text := "By induction and then \cref{lem:zero-add, lem:succ-add}." }
-- Miscellaneous metadata
notReady := false, discussion := none, title := none
\end{lstlisting}

\subsection{\LaTeX{} Export} \label{sec:latex-export}

To build the \LaTeX{} blueprint from such data, LeanArchitect provides a build-time export mechanism integrated with Lean's Lake build system, inspired by doc-gen4. For each module, blueprint data is extracted and rendered into \LaTeX{} fragments that can be imported into an existing blueprint document using the macro \bplatexcode|\inputleannode{label}|. Multiple Lean declarations are allowed to correspond to a single blueprint node. The export process is deterministic and incremental. For example, when the user writes \bplatexcode|\inputleannode{thm:add-comm}| in the \LaTeX{} blueprint, it will be expanded to:

\begin{lstlisting}[
  language={[latex]tex},
  morecomment={[l][\color{commentcolor}]{\%}},
  commentstyle={\ttfamily\footnotesize},
  moretexcs={lean,leanok,uses,cref}
]
% Statement
\begin{theorem}
  \label{thm:add-comm} \lean{MyNat.add_comm}  % Identifiers
  \leanok \uses{def:nat}                      % Status and dependencies
  Addition in $ℕ$ is commutative.             % Text
\end{theorem}

% Proof
\begin{proof}
  \uses{lem:zero-add, lem:succ-add}           % Status and dependencies
  By induction and then                       % Text
  \cref{lem:zero-add, lem:succ-add}.
\end{proof}
\end{lstlisting}

Then, the final step to produce a PDF and web visualization of the blueprint and its dependency graph is done using \leanblueprint.

\subsection{Project Management Using \leanarchitect}
We envision the workflow of a large Lean project using \leanarchitect to be as follows.

\begin{enumerate}
\item Mathematicians write a detailed exposition of the formalization target in a \LaTeX{} document (e.g.\ on Overleaf), following some source material.
\item A new formalization project is set up (e.g.\ on GitHub) with the document as a blueprint.
\item \label{itm:tag-blueprint} Lean experts translate each theorem or definition into Lean, tagging each one with \leancode{@[blueprint]}. This translation can leave \leancode{sorry} placeholders in the code.
\item \label{itm:replace-inputleannode} In the \LaTeX{} blueprint, managers can then use \bplatexcode{\inputleannode} to replace existing theorems with automatically inferred blueprint metadata.
\item Lean experts fill the \leancode{sorry}s in the code, while \leanarchitect automatically updates the metadata.
\end{enumerate}

We also provide \leancode{blueprintConvert}, a script that replaces a theorem or definition in the \LaTeX{} blueprint with \bplatexcode{\inputleannode}, and inserts an appropriate \leancode{@[blueprint]} tag into the Lean code. This script is primarily used to automatically migrate existing \leanblueprint-based projects to \leanarchitect format (see~\cref{sec:case-study-convert}), but it can also automate steps \ref{itm:tag-blueprint}--\ref{itm:replace-inputleannode} above.

\section{Case Studies}

\subsection{PrimeNumberTheoremAnd}

PrimeNumberTheoremAnd~\cite{kontorovich_pnt} is a large community-driven project aiming to formalize the prime number theorem and related results in analytic number theory in Lean. It is the first external project to adopt \leanarchitect as its primary blueprint infrastructure. The project began in 2023 and was migrated to \leanarchitect in January 2026.

Prior to migration, PrimeNumberTheoremAnd used a custom tool (\emph{leanblueprint-extract}) that embedded blueprint information as specially formatted comments inside Lean files and extracted them into \LaTeX{} for blueprint generation. This workflow enabled co-locating formal code and informal exposition, but required manual maintenance of metadata such as \latexcode{\leanok} and \latexcode{\uses}. (The design of \leanarchitect, in particular the \bplatexcode{\inputleanmodule} mechanism, was inspired by this approach.)

Migration to \leanarchitect required only surface-level syntactic changes, primarily replacing custom comments with \leancode{@[blueprint]} attributes and \leancode{blueprint_comment} commands.\footnote{See the \href{https://github.com/AlexKontorovich/PrimeNumberTheoremAnd/pull/439}{GitHub pull request}.} The initial conversion took approximately one day of work by us. After conversion, blueprint metadata---including \latexcode{\lean}, \latexcode{\leanok}, \latexcode{\uses}, and \latexcode{\mathlibok}---is now inferred automatically from Lean, eliminating a significant source of duplication and reducing maintenance overhead.

A minor source of disruption was that existing pull requests had to be manually updated by their authors to conform to the new annotation format. From the project maintainer's perspective, the impact was immediately positive. In particular, the automatic coloring of nodes and management of \latexcode{\leanok} tags substantially improved the usability of the blueprint as a progress-tracking and coordination tool. As Terence Tao commented publicly on Zulip:\footnote{See the \href{https://leanprover.zulipchat.com/\#narrow/channel/423402-PrimeNumberTheorem.2B/topic/Update.20on.20tertiary.20estimate.20projects/near/567004857}{Zulip comment}.}
\begin{quote}
The auto-coloring of the blueprint and management of the \textbackslash{}leanok tags is very pleasant from the project maintainer side of things!
\end{quote}

Another advantage of \leanarchitect is \emph{dependency debugging}. Tao observed that although the statement of a lemma (Erd\H{o}s 392) was syntactically correct, its statement was semantically incorrect, as its AI-generated proof did not depend on surrounding lemmas. This problem was surfaced by the visualization of dependency relations automatically inferred by \leanarchitect.

Overall, this case study demonstrates that \leanarchitect can be adopted in an active, large-scale project, preserve existing authoring workflows, and improve synchronization and maintainability of blueprint metadata with small migration cost.

\subsection{Converting Existing Projects to \leanarchitect Format}
\label{sec:case-study-convert}

We tested the conversion script on the following projects into \leanarchitect, at Lean version~\bashcode{v4.25.0}:
\begin{enumerate}
\item Carleson~\cite{fvp_carleson}
\item Brownian Motion~\cite{degenne2025brownian}
\item Infinity Cosmos~\cite{riehl_infinity-cosmos}
\item Fermat's Last Theorem~\cite{buzzard_flt}
\item Prime Number Theorem And~\cite{kontorovich_pnt}
\end{enumerate}

For each project, we first added \leanarchitect to the lakefile, and then ran the conversion script. For the most part, the script is able to automatically convert blueprint nodes into Lean \leancode|@[blueprint]| attributes.

During conversion, we found some discrepancies between the converted blueprint and the original one. This revealed some issues with the original manually written blueprint that could be uncovered and automatically fixed by \leanarchitect, such as:
\begin{itemize}
  \item Isolated nodes that were actually not used, such as a layer-cake theorem \leancode{eLpNorm_pow_eq_distribution} not used in Carleson (a related theorem \leancode{eLpNorm_eq_distribution} was used instead)
  \item Missing dependency edges, such as \leancode{internalCoveringNumber_eq_one_of_diam_le} incorrectly marked as unused in Brownian Motion
  \item Theorems in Mathlib that should have been marked \latexcode{\mathlibok}
  \item Oversimplified setups, such as Infinity Cosmos only applying \latexcode{\leanok} and \latexcode{\uses} on statements and not proofs of theorems.
\end{itemize}

In general, the conversion script successfully converts projects into \leanarchitect format. As \leanarchitect automates the previously manual specification of \latexcode{\leanok} and \latexcode{\uses}, it ensures the blueprint is in sync with the Lean formalization status.

\subsection{Blueprint-Based Autoformalization}


We evaluate LeanArchitect as an interface for human--AI collaboration by semi-automatically formalizing a theorem that is both technically nontrivial and naturally decomposable: multivariate Taylor's theorem in integral form. At the time of writing, Mathlib contains Taylor's theorem only in one-dimensional settings and does not have the multivariate Fr\'echet-derivative formulation in integral form.

\paragraph{Target theorem}
Let $E, F$ be normed vector spaces over $\mathbb{R}$ with $F$ complete. Let $U \subseteq E$ be open and let $f : U \to F$ be $C^n$ with $n>0$. For $x,h \in E$ such that $[x,x+h] \subseteq U$, we aim to formalize:
\begin{align*}
f(x+h) ={}& \sum_{k=0}^{n-1}\frac{1}{k!}\, D^k f(x)[h,\dots,h]
+ \frac{1}{(n-1)!}\int_0^1(1-s)^{n-1}\,D^n f(x+sh)[h,\dots,h]\,\mathrm{d}s,
\end{align*}
where $D^k f$ denotes the $k$-th Fr\'echet derivative.

\paragraph{Pipeline}
Our pipeline uses two AI systems with complementary roles: GPT-5 Pro drafts a natural-language proof and a corresponding Lean \emph{blueprint file} (a single Lean module annotated with \leancode{@[blueprint]}), and Aristotle operates in \emph{sorry-filling} mode to attempt to complete the resulting proof obligations.
Concretely:
\begin{enumerate}
\item GPT-5 Pro produces a structured proof outline and translates it into a Lean file whose intermediate lemmas are tagged with \leancode{@[blueprint]} and initially proved with \leancode{sorry}.
\item We visualize progress using the generated blueprint dependency graph.
\item Aristotle (sorry-filling mode) attempts to discharge the \leancode{sorry} nodes automatically.
\item Remaining failures are iteratively addressed by refining the blueprint using GPT-5 Pro.
\item Finally, a human manually proves remaining parts that could not be automatically proved.
\end{enumerate}

\begin{figure}[t]
  \centering
  \begin{subfigure}{0.48\linewidth}
  \centering
  \includegraphics[width=\linewidth]{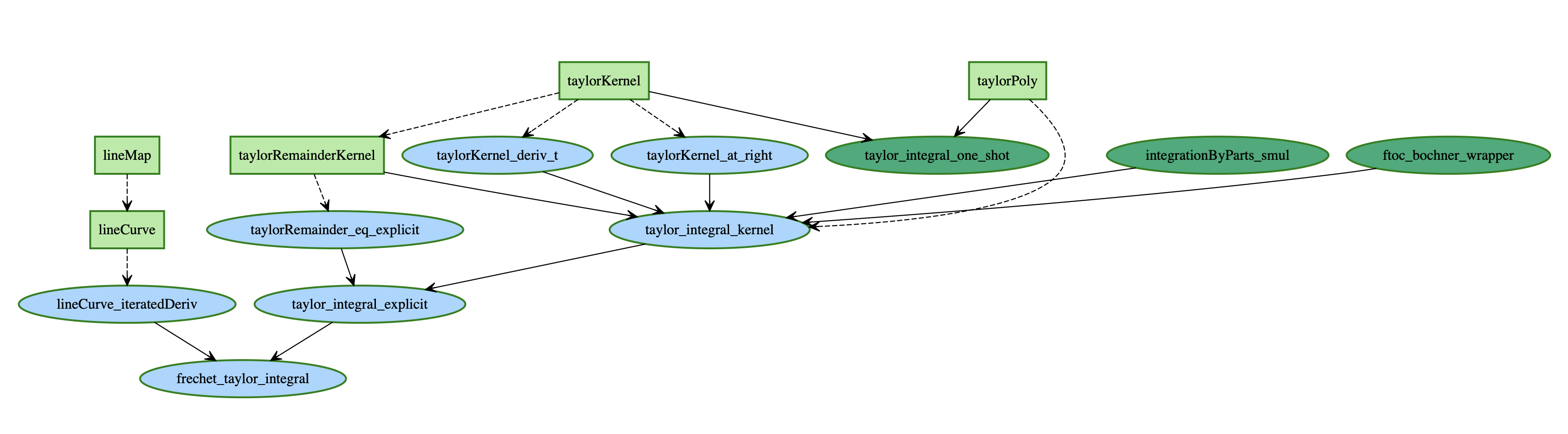}
  \caption{GPT-5 Pro: initial blueprint.} \label{fig:taylor-gpt-round1}
  \end{subfigure}\hfill
  \begin{subfigure}{0.48\linewidth}
  \centering
  \includegraphics[width=\linewidth]{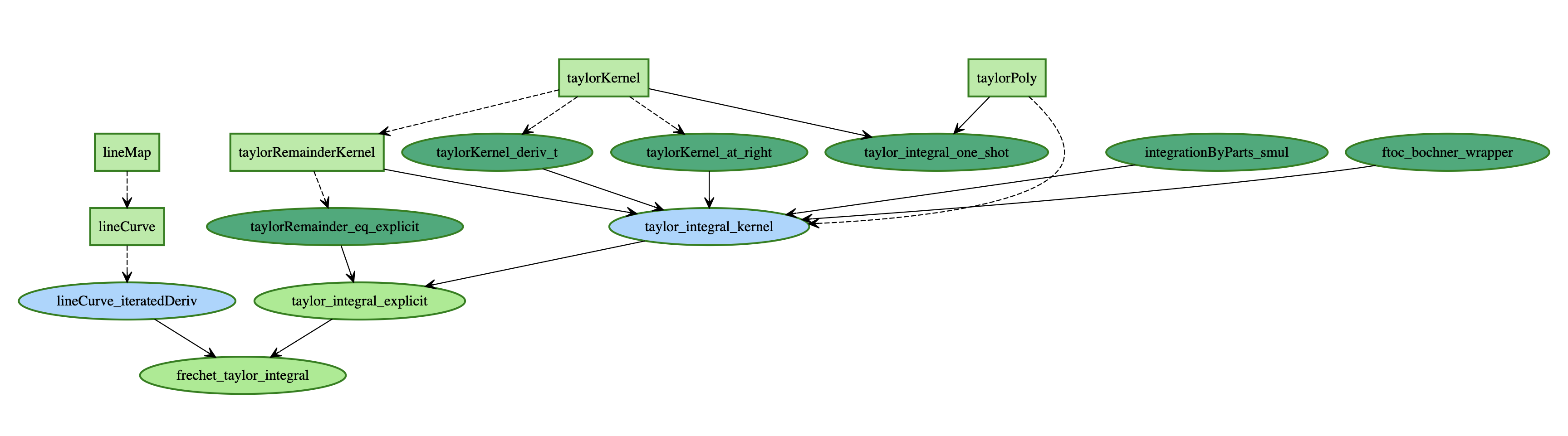}
  \caption{Aristotle: fill proofs in initial blueprint.} \label{fig:taylor-aristotle-round1}
  \end{subfigure}

  \medskip

  \begin{subfigure}{0.48\linewidth}
  \centering
  \includegraphics[width=\linewidth]{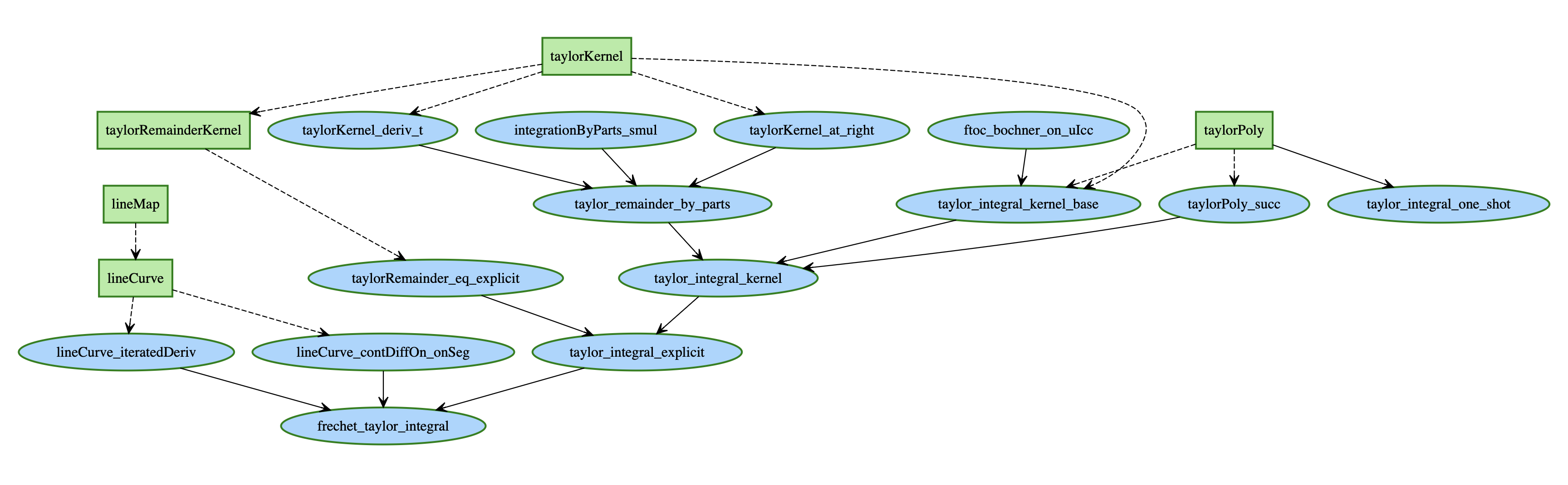}
  \caption{GPT-5 Pro: refined blueprint.} \label{fig:taylor-gpt-round2}
  \end{subfigure}\hfill
  \begin{subfigure}{0.48\linewidth}
  \centering
  \includegraphics[width=\linewidth]{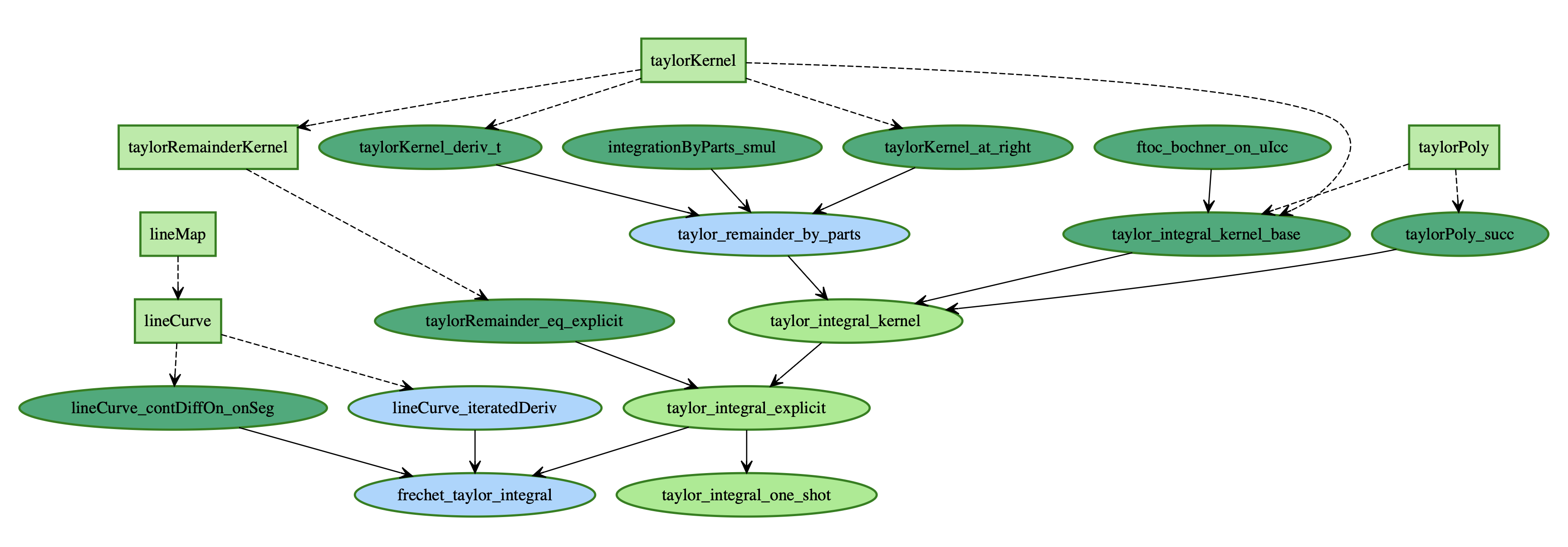}
  \caption{Aristotle: fill proofs in refined blueprint.} \label{fig:taylor-aristotle-round2}
  \end{subfigure}

  \medskip
  
  \begin{subfigure}{0.48\linewidth}
  \centering
  \includegraphics[width=\linewidth]{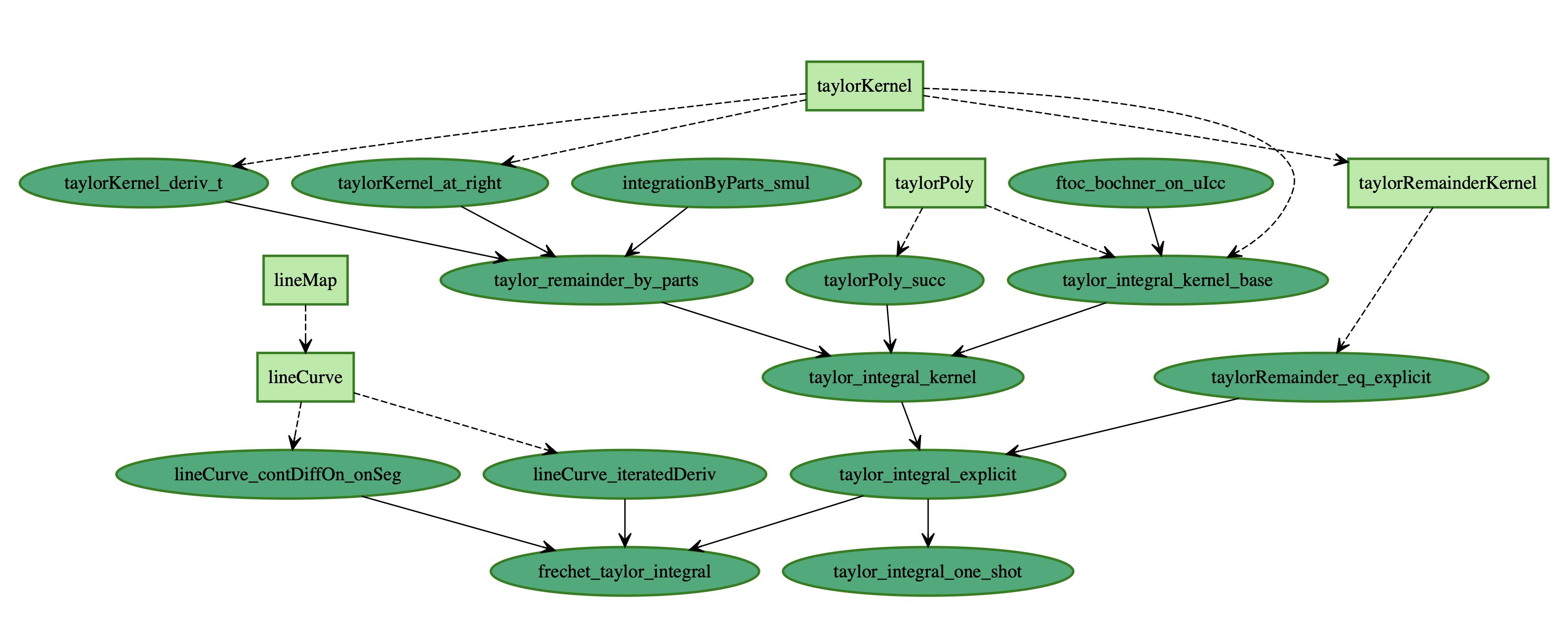}
  \caption{Completed formalization by human.} \label{fig:taylor-complete}
  \end{subfigure}

  \caption{Progress of blueprint-based autoformalization for multivariate Taylor's theorem.
  Blue nodes are unproved lemmas; green nodes are proved lemmas. Rectangles denote
  definitions; circles denote theorems.}
  \label{fig:taylor_blueprint_progress}
\end{figure}

\paragraph{Progress visualization}
\Cref{fig:taylor-gpt-round1} shows the initial blueprint produced from GPT-5 Pro's Lean draft, and \Cref{fig:taylor-aristotle-round1} shows the result after Aristotle's first pass. We then refined the blueprint based on the remaining unproved nodes (\Cref{fig:taylor-gpt-round2,fig:taylor-aristotle-round2}). After refinement, Aristotle proved all but three nodes.

Inspection of the remaining failures revealed a systematic issue in the initial formalization: one should use \leancode{derivWithin} rather than \leancode{deriv} for a function $f\colon\mathbb{R}\to E$ differentiable on~$[a,b]$. We corrected the affected statements manually, making intermediate lemmas true, and then proved the remaining lemmas by hand.

\begin{samepage}
\paragraph{Outcome}
This proves the target theorem in Lean:
\begin{lstlisting}
import Mathlib

open scoped Nat

variable {E : Type*} [NormedAddCommGroup E] [NormedSpace ℝ E]
variable {F : Type*} [NormedAddCommGroup F] [NormedSpace ℝ F] [CompleteSpace F]

theorem §\declname{frechet\_taylor\_integral}§
  {U : Set E} {f : E → F} {n : ℕ} (hn : 0 < n)
  (hCn : ContDiffOn ℝ n f U) (hU : IsOpen U)
  {x h : E} (hseg : ∀ s ∈ Set.Icc (0 : ℝ) 1, x + s • h ∈ U) :
  f (x + h) =
  (∑ k ∈ Finset.range n, ((k)! : ℝ)⁻¹ • iteratedFDeriv ℝ k f x fun _ ↦ h) +
  ((n - 1)! : ℝ)⁻¹ • ∫ s in (0 : ℝ)..1,
    (1 - s) ^ (n - 1) • iteratedFDeriv ℝ n f (x + s • h) fun _ ↦ h
\end{lstlisting}
\end{samepage}

This case study highlights two advantages of blueprint-structured autoformalization. First, failures are localized: rather than treating the attempt as an all-or-nothing proof search, we can focus human attention on a small set of remaining nodes identified in the dependency graph. Second, progress is transparent and compositional: partial success yields reusable intermediate lemmas, and the dependency structure provides a natural unit of work for automated provers. More broadly, the blueprint acts as a shared interface between humans and automation: humans refine the decomposition and hypotheses, while automated systems attempt the resulting subgoals.

We also observed a limitation of the current human--AI loop: simply asking the language model to ``refine'' the blueprint did not reliably address the underlying causes of failure, partly because the prover provides limited diagnostic feedback beyond a list of unsolved nodes. Closing this loop will likely require richer feedback signals (e.g.\ counterexamples, missing lemma suggestions, or structured failure traces) that can be consumed by the planner model.

\section{Limitations}

LeanArchitect introduces a new integration point between Lean and blueprint workflows and has several limitations. First, authoring \LaTeX{} inside Lean files is currently unergonomic: mainstream Lean IDE support provides neither syntax highlighting nor interactive feedback for embedded \LaTeX{} in \bashcode{.lean} files, and existing keybindings interfere with natural \LaTeX{} input. Second, LeanArchitect supports a many-to-one correspondence from Lean declarations to blueprint nodes: multiple Lean declarations may correspond to the same blueprint node, but a given Lean declaration cannot correspond to more than one blueprint node, which limits expressiveness in cases where the same lemma appears in the blueprint multiple times. Third, the additional build steps of \leanarchitect deepen the build pipeline and increase the possibility for build and CI failures. Finally, blueprint nodes that exist only in the informal \LaTeX{} layer must still be managed manually until a corresponding Lean declaration is introduced. These limitations primarily reflect tooling and infrastructure gaps rather than fundamental design constraints.

\section*{Acknowledgements}
Work partially supported by NSF Grant DMS-2434614 and a gift from Convergent Research.

\bibliographystyle{plainnat}
\bibliography{refs}

\begin{thebibliography}{26}
\providecommand{\natexlab}[1]{#1}
\providecommand{\url}[1]{\texttt{#1}}
\expandafter\ifx\csname urlstyle\endcsname\relax
  \providecommand{\doi}[1]{doi: #1}\else
  \providecommand{\doi}{doi: \begingroup \urlstyle{rm}\Url}\fi

\bibitem[Achim et~al.(2025)Achim, Best, Bietti, Der, F{\'e}d{\'e}rico, Gukov, Halpern-Leistner, Henningsgard, Kudryashov, Meiburg, et~al.]{achim2025aristotle}
Tudor Achim, Alex Best, Alberto Bietti, Kevin Der, Math{\"i}s F{\'e}d{\'e}rico, Sergei Gukov, Daniel Halpern-Leistner, Kirsten Henningsgard, Yury Kudryashov, Alexander Meiburg, et~al.
\newblock {Aristotle: IMO-Level Automated Theorem Proving}.
\newblock \emph{arXiv preprint}, 2025.
\newblock \doi{10.48550/arXiv.2510.01346}.
\newblock URL \url{https://doi.org/10.48550/arXiv.2510.01346}.

\bibitem[Becker et~al.(2025)Becker, de~Frutos-Fern{\'a}ndez, Diedering, van Doorn, Gou{\"e}zel, Jamneshan, Karunus, van~de Meent, Monticone, Mulder-Sohn, et~al.]{fvp_carleson}
Lars Becker, Mar{\'i}a~In{\'e}s de~Frutos-Fern{\'a}ndez, Leo Diedering, Floris van Doorn, S{\'e}bastien Gou{\"e}zel, Asgar Jamneshan, Evgenia Karunus, Edward van~de Meent, Pietro Monticone, Jasper Mulder-Sohn, et~al.
\newblock {A Blueprint for the Formalization of Carleson's Theorem on Convergence of Fourier Series}.
\newblock \emph{arXiv preprint}, 2025.
\newblock \doi{10.48550/arXiv.2405.06423}.
\newblock URL \url{https://doi.org/10.48550/arXiv.2405.06423}.

\bibitem[Bolan et~al.(2025)Bolan, Breitner, Brox, Carlini, Carneiro, van Doorn, Dvorak, Goens, Hill, Husum, Ibarra~Mejia, Kocsis, Le~Floch, Livne Bar-on, Luccioli, McNeil, Meiburg, Monticone, Nielsen, Osazuwa, Paolini, Petracci, Reinke, Renshaw, Rossel, Roux, Scanvic, Srinivas, Tadipatri, Tao, Tsyrklevich, Vaquerizo-Villar, Weber, and Zheng]{tao2025equational}
Matthew Bolan, Joachim Breitner, Jose Brox, Nicholas Carlini, Mario Carneiro, Floris van Doorn, Martin Dvorak, Andr{\'e}s Goens, Aaron Hill, Harald Husum, Hern{\'a}n Ibarra~Mejia, Zoltan~A. Kocsis, Bruno Le~Floch, Amir Livne Bar-on, Lorenzo Luccioli, Douglas McNeil, Alex Meiburg, Pietro Monticone, Pace~P. Nielsen, Emmanuel~Osalotioman Osazuwa, Giovanni Paolini, Marco Petracci, Bernhard Reinke, David Renshaw, Marcus Rossel, Cody Roux, J{\'e}r{\'e}my Scanvic, Shreyas Srinivas, Anand~Rao Tadipatri, Terence Tao, Vlad Tsyrklevich, Fernando Vaquerizo-Villar, Daniel Weber, and Fan Zheng.
\newblock {The Equational Theories Project: Advancing Collaborative Mathematical Research at Scale}.
\newblock \emph{arXiv preprint}, 2025.
\newblock \doi{10.48550/arXiv.2512.07087}.
\newblock URL \url{https://doi.org/10.48550/arXiv.2512.07087}.

\bibitem[Buzzard and Taylor(2023)]{buzzard_flt}
Kevin Buzzard and Richard Taylor.
\newblock {FLT: An Ongoing Lean Formalisation of the Proof of Fermat's Last Theorem}, 2023.
\newblock URL \url{https://github.com/ImperialCollegeLondon/FLT}.

\bibitem[Chen et~al.(2025{\natexlab{a}})Chen, Chen, Du, Hu, Jiang, Jie, Jin, Jin, Li, Shi, et~al.]{chen2025seed1.5}
Jiangjie Chen, Wenxiang Chen, Jiacheng Du, Jinyi Hu, Zhicheng Jiang, Allan Jie, Xiaoran Jin, Xing Jin, Chenggang Li, Wenlei Shi, et~al.
\newblock {Seed-Prover 1.5: Mastering Undergraduate-Level Theorem Proving via Learning from Experience}.
\newblock \emph{arXiv preprint}, 2025{\natexlab{a}}.
\newblock \doi{10.48550/arXiv.2512.17260}.
\newblock URL \url{https://doi.org/10.48550/arXiv.2512.17260}.

\bibitem[Chen et~al.(2025{\natexlab{b}})Chen, Gu, Huang, Huang, Jiang, Jie, Jin, Jin, Li, Ma, et~al.]{chen2025seed}
Luoxin Chen, Jinming Gu, Liankai Huang, Wenhao Huang, Zhicheng Jiang, Allan Jie, Xiaoran Jin, Xing Jin, Chenggang Li, Kaijing Ma, et~al.
\newblock {Seed-Prover: Deep and Broad Reasoning for Automated Theorem Proving}.
\newblock \emph{arXiv preprint}, 2025{\natexlab{b}}.
\newblock \doi{10.48550/arXiv.2507.23726}.
\newblock URL \url{https://doi.org/10.48550/arXiv.2507.23726}.

\bibitem[de~Moura and Ullrich(2021)]{demoura2021lean}
Leonardo de~Moura and Sebastian Ullrich.
\newblock {The Lean 4 Theorem Prover and Programming Language}.
\newblock In \emph{Automated Deduction -- {CADE} 28}, volume 12699 of \emph{Lecture Notes in Computer Science}, pages 625--635. Springer, 2021.
\newblock \doi{10.1007/978-3-030-79876-5_37}.
\newblock URL \url{https://doi.org/10.1007/978-3-030-79876-5_37}.

\bibitem[de~Moura et~al.(2015)de~Moura, Kong, Avigad, van Doorn, and von Raumer]{demoura2015lean}
Leonardo de~Moura, Soonho Kong, Jeremy Avigad, Floris van Doorn, and Jakob von Raumer.
\newblock {The Lean Theorem Prover (System Description)}.
\newblock In \emph{Automated Deduction -- {CADE-25}}, volume 9195 of \emph{Lecture Notes in Computer Science}, pages 378--388. Springer, 2015.
\newblock \doi{10.1007/978-3-319-21401-6_26}.
\newblock URL \url{https://doi.org/10.1007/978-3-319-21401-6_26}.

\bibitem[Degenne et~al.(2025)Degenne, Ledvinka, Marion, and Pfaffelhuber]{degenne2025brownian}
R{\'e}my Degenne, David Ledvinka, Etienne Marion, and Peter Pfaffelhuber.
\newblock {Formalization of Brownian Motion in Lean}.
\newblock \emph{arXiv preprint}, 2025.
\newblock \doi{10.48550/arXiv.2511.20118}.
\newblock URL \url{https://doi.org/10.48550/arXiv.2511.20118}.

\bibitem[{Google DeepMind}(2025)]{deepmind_formalconjectures}
{Google DeepMind}.
\newblock {Formal Conjectures: A Collection of Formalized Statements of Conjectures in Lean}, 2025.
\newblock URL \url{https://github.com/google-deepmind/formal-conjectures}.

\bibitem[Hu et~al.(2025)Hu, Zhu, and Welleck]{hu2024minictx}
Jiewen Hu, Thomas Zhu, and Sean Welleck.
\newblock {miniCTX: Neural Theorem Proving with (Long-)Contexts}.
\newblock In \emph{The Thirteenth International Conference on Learning Representations}, 2025.
\newblock URL \url{https://openreview.net/forum?id=KIgaAqEFHW}.

\bibitem[Hubert et~al.(2025)Hubert, Mehta, Sartran, Horv{\'a}th, {\v{Z}}u{\v{z}}i{\'c}, Wieser, Huang, Schrittwieser, Schroecker, Masoom, et~al.]{hubert2025olympiad}
Thomas Hubert, Rishi Mehta, Laurent Sartran, Mikl{\'o}s~Z. Horv{\'a}th, Goran {\v{Z}}u{\v{z}}i{\'c}, Eric Wieser, Aja Huang, Julian Schrittwieser, Yannick Schroecker, Hussain Masoom, et~al.
\newblock {Olympiad-Level Formal Mathematical Reasoning with Reinforcement Learning}.
\newblock \emph{Nature}, 2025.
\newblock \doi{10.1038/s41586-025-09833-y}.
\newblock URL \url{https://doi.org/10.1038/s41586-025-09833-y}.

\bibitem[Kontorovich and Tao(2024)]{kontorovich_pnt}
Alex Kontorovich and Terence Tao.
\newblock {Prime Number Theorem and More}, 2024.
\newblock URL \url{https://github.com/AlexKontorovich/PrimeNumberTheoremAnd}.

\bibitem[Lin et~al.(2025{\natexlab{a}})Lin, Tang, Lyu, Wu, Lin, Yang, Li, Xia, Chen, Arora, et~al.]{lin2025goedel}
Yong Lin, Shange Tang, Bohan Lyu, Jiayun Wu, Hongzhou Lin, Kaiyu Yang, Jia Li, Mengzhou Xia, Danqi Chen, Sanjeev Arora, et~al.
\newblock {Goedel-Prover: A Frontier Model for Open-Source Automated Theorem Proving}.
\newblock \emph{arXiv preprint}, 2025{\natexlab{a}}.
\newblock \doi{10.48550/arXiv.2502.07640}.
\newblock URL \url{https://doi.org/10.48550/arXiv.2502.07640}.

\bibitem[Lin et~al.(2025{\natexlab{b}})Lin, Tang, Lyu, Yang, Chung, Zhao, Jiang, Geng, Ge, Sun, et~al.]{lin2025goedel2}
Yong Lin, Shange Tang, Bohan Lyu, Ziran Yang, Jui-Hui Chung, Haoyu Zhao, Lai Jiang, Yihan Geng, Jiawei Ge, Jingruo Sun, et~al.
\newblock {Goedel-Prover-V2: Scaling Formal Theorem Proving with Scaffolded Data Synthesis and Self-Correction}.
\newblock \emph{arXiv preprint}, 2025{\natexlab{b}}.
\newblock \doi{10.48550/arXiv.2508.03613}.
\newblock URL \url{https://doi.org/10.48550/arXiv.2508.03613}.

\bibitem[Massot(2020)]{massot_leanblueprint}
Patrick Massot.
\newblock {leanblueprint: plasTeX Plugin to Build Formalization Blueprints}, 2020.
\newblock URL \url{https://github.com/PatrickMassot/leanblueprint}.

\bibitem[Poiroux et~al.(2025)Poiroux, Bosselut, and Kun{\v{c}}ak]{poiroux2025rlmeval}
Auguste Poiroux, Antoine Bosselut, and Viktor Kun{\v{c}}ak.
\newblock {RLMEval: Evaluating Research-Level Neural Theorem Proving}.
\newblock In \emph{Findings of the Association for Computational Linguistics: {EMNLP} 2025}, 2025.
\newblock \doi{10.48550/arXiv.2510.25427}.
\newblock URL \url{https://doi.org/10.48550/arXiv.2510.25427}.

\bibitem[Ren et~al.(2025)Ren, Shao, Song, Xin, Wang, Zhao, Zhang, Fu, Zhu, Yang, et~al.]{ren2025deepseek}
Z.~Z. Ren, Zhihong Shao, Junxiao Song, Huajian Xin, Haocheng Wang, Wanjia Zhao, Liyue Zhang, Zhe Fu, Qihao Zhu, Dejian Yang, et~al.
\newblock {DeepSeek-Prover-V2: Advancing Formal Mathematical Reasoning via Reinforcement Learning for Subgoal Decomposition}.
\newblock \emph{arXiv preprint}, 2025.
\newblock \doi{10.48550/arXiv.2504.21801}.
\newblock URL \url{https://doi.org/10.48550/arXiv.2504.21801}.

\bibitem[Riehl and Verity(2024)]{riehl_infinity-cosmos}
Emily Riehl and Dominic Verity.
\newblock {Infinity Cosmos: A Blueprint for a Formalization of Infinity-Cosmos Theory in Lean}, 2024.
\newblock URL \url{https://github.com/emilyriehl/infinity-cosmos}.

\bibitem[{The mathlib Community}(2020)]{The_mathlib_Community_2020}
{The mathlib Community}.
\newblock {The Lean Mathematical Library}.
\newblock In \emph{Proceedings of the 9th {ACM} {SIGPLAN} International Conference on Certified Programs and Proofs}, {CPP} '20, pages 367--381. ACM, 2020.
\newblock \doi{10.1145/3372885.3373824}.
\newblock URL \url{https://doi.org/10.1145/3372885.3373824}.

\bibitem[Wang et~al.(2025)Wang, Unsal, Lin, Baksys, Liu, Santos, Sung, Vinyes, Ying, Zhu, et~al.]{wang2025kimina}
Haiming Wang, Mert Unsal, Xiaohan Lin, Mantas Baksys, Junqi Liu, Marco~Dos Santos, Flood Sung, Marina Vinyes, Zhenzhe Ying, Zekai Zhu, et~al.
\newblock {Kimina-Prover Preview: Towards Large Formal Reasoning Models with Reinforcement Learning}.
\newblock \emph{arXiv preprint}, 2025.
\newblock \doi{10.48550/arXiv.2504.11354}.
\newblock URL \url{https://doi.org/10.48550/arXiv.2504.11354}.

\bibitem[Xin et~al.(2024{\natexlab{a}})Xin, Guo, Shao, Ren, Zhu, Liu, Ruan, Li, and Liang]{xin2024deepseek}
Huajian Xin, Daya Guo, Zhihong Shao, Zhizhou Ren, Qihao Zhu, Bo~Liu, Chong Ruan, Wenda Li, and Xiaodan Liang.
\newblock {DeepSeek-Prover: Advancing Theorem Proving in LLMs through Large-Scale Synthetic Data}.
\newblock \emph{arXiv preprint}, 2024{\natexlab{a}}.
\newblock \doi{10.48550/arXiv.2405.14333}.
\newblock URL \url{https://doi.org/10.48550/arXiv.2405.14333}.

\bibitem[Xin et~al.(2024{\natexlab{b}})Xin, Ren, Song, Shao, Zhao, Wang, Liu, Zhang, Lu, Du, et~al.]{xin2024deepseek1.5}
Huajian Xin, Z.~Z. Ren, Junxiao Song, Zhihong Shao, Wanjia Zhao, Haocheng Wang, Bo~Liu, Liyue Zhang, Xuan Lu, Qiushi Du, et~al.
\newblock {DeepSeek-Prover-V1.5: Harnessing Proof Assistant Feedback for Reinforcement Learning and Monte-Carlo Tree Search}.
\newblock \emph{arXiv preprint}, 2024{\natexlab{b}}.
\newblock \doi{10.48550/arXiv.2408.08152}.
\newblock URL \url{https://doi.org/10.48550/arXiv.2408.08152}.

\bibitem[Yang et~al.(2023)Yang, Swope, Gu, Chalamala, Song, Yu, Godil, Prenger, and Anandkumar]{yang2023leandojo}
Kaiyu Yang, Aidan Swope, Alex Gu, Rahul Chalamala, Peiyang Song, Shixing Yu, Saad Godil, Ryan~J. Prenger, and Animashree Anandkumar.
\newblock {LeanDojo: Theorem Proving with Retrieval-Augmented Language Models}.
\newblock \emph{{Advances in Neural Information Processing Systems}}, 36:\penalty0 21573--21612, 2023.

\bibitem[Zheng et~al.(2021)Zheng, Han, and Polu]{zheng2021minif2f}
Kunhao Zheng, Jesse~Michael Han, and Stanislas Polu.
\newblock {miniF2F: A Cross-System Benchmark for Formal Olympiad-Level Mathematics}.
\newblock \emph{arXiv preprint}, 2021.
\newblock \doi{10.48550/arXiv.2109.00110}.
\newblock URL \url{https://doi.org/10.48550/arXiv.2109.00110}.

\bibitem[Zhu et~al.(2025)Zhu, Clune, Avigad, Jiang, and Welleck]{zhu2025premise}
Thomas Zhu, Joshua Clune, Jeremy Avigad, Albert~Qiaochu Jiang, and Sean Welleck.
\newblock {Premise Selection for a Lean Hammer}.
\newblock \emph{arXiv preprint}, 2025.
\newblock \doi{10.48550/arXiv.2506.07477}.
\newblock URL \url{https://doi.org/10.48550/arXiv.2506.07477}.

\end{thebibliography}

\appendix
\section{Technical Details}

\subsection{Example}
\label{lean-example}

We begin with an illustrative example of \leanarchitect.
\begin{lstlisting}
import Architect

@[blueprint]
inductive §\declname{MyNat}§ : Type where
  | zero : MyNat
  | succ : MyNat → MyNat

namespace MyNat

@[blueprint "def:nat-add"
  (statement := /-- Natural number addition. -/)]
def §\declname{add}§ (a b : MyNat) : MyNat :=
  match b with
  | zero => a
  | succ b => succ (add a b)

@[simp, blueprint
  (statement := /-- For any natural number $a$, $0 + a = a$,
  where $+$ is \cref{def:nat-add}. -/)]
theorem §\declname{zero\_add}§ (a : MyNat) : add zero a = a := by
  /-- The proof follows by induction. -/
  induction a <;> simp [*, add]

@[blueprint
  (statement := /-- For any natural numbers $a, b$,
  $(a + 1) + b = (a + b) + 1$. -/)]
theorem §\declname{succ\_add}§ (a b : MyNat) : add (succ a) b = succ (add a b) := by
  /-- Proof by induction on $b$. -/
  sorry

@[blueprint
  (statement := /-- For any natural numbers $a, b$,
  $a + b = b + a$. -/)]
theorem §\declname{add\_comm}§ (a b : MyNat) : add a b = add b a := by
  induction b with
  | zero =>
  have := trivial
  /-- The base case follows from \cref{MyNat.zero_add}. -/
  simp [add]
  | succ b ih =>
  /-- The inductive case follows from \cref{MyNat.succ_add}. -/
  sorry_using [succ_add]  -- the `sorry_using` tactic declares dependency

-- Additional content omitted

end MyNat
\end{lstlisting}

\begin{figure}[ht]
  \centering
  \begin{subfigure}[t]{0.4\textwidth}
      \centering
      \fbox{\includegraphics[height=2in]{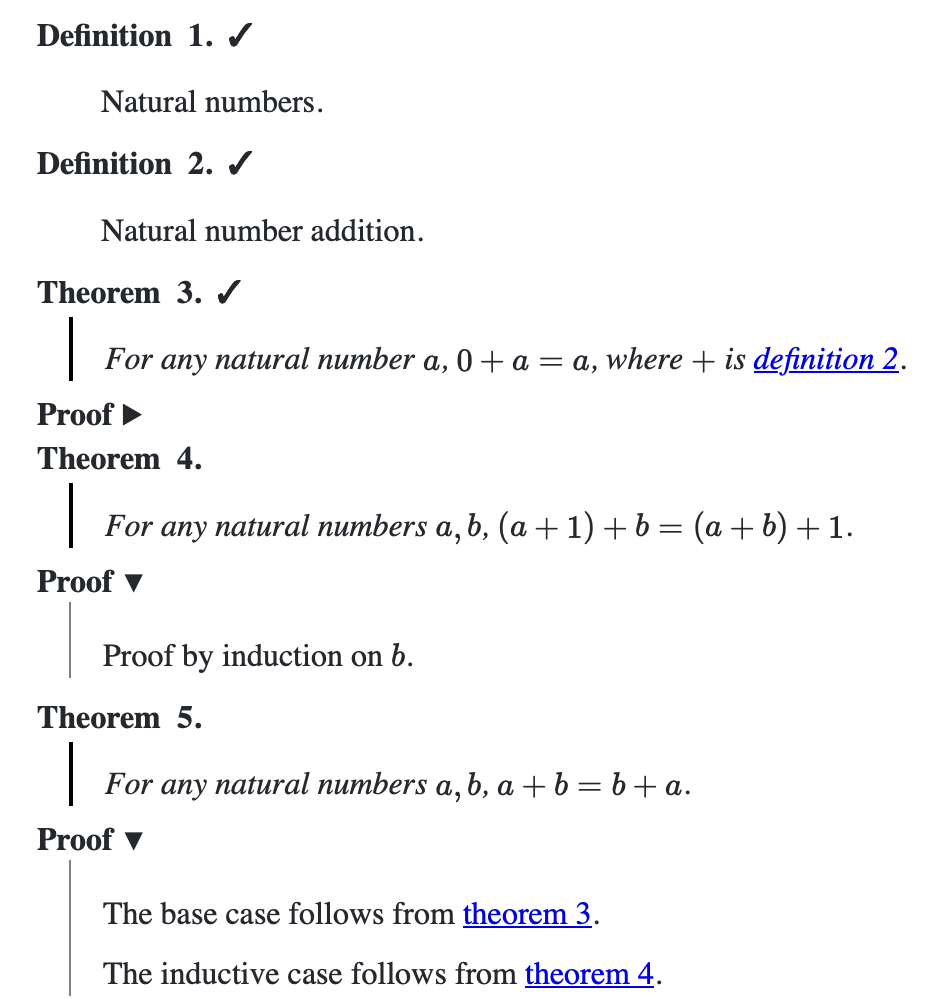}}
      \caption{Blueprint document.}
  \end{subfigure}%
  \begin{subfigure}[t]{0.6\textwidth}
      \centering
      \includegraphics[width=1\linewidth]{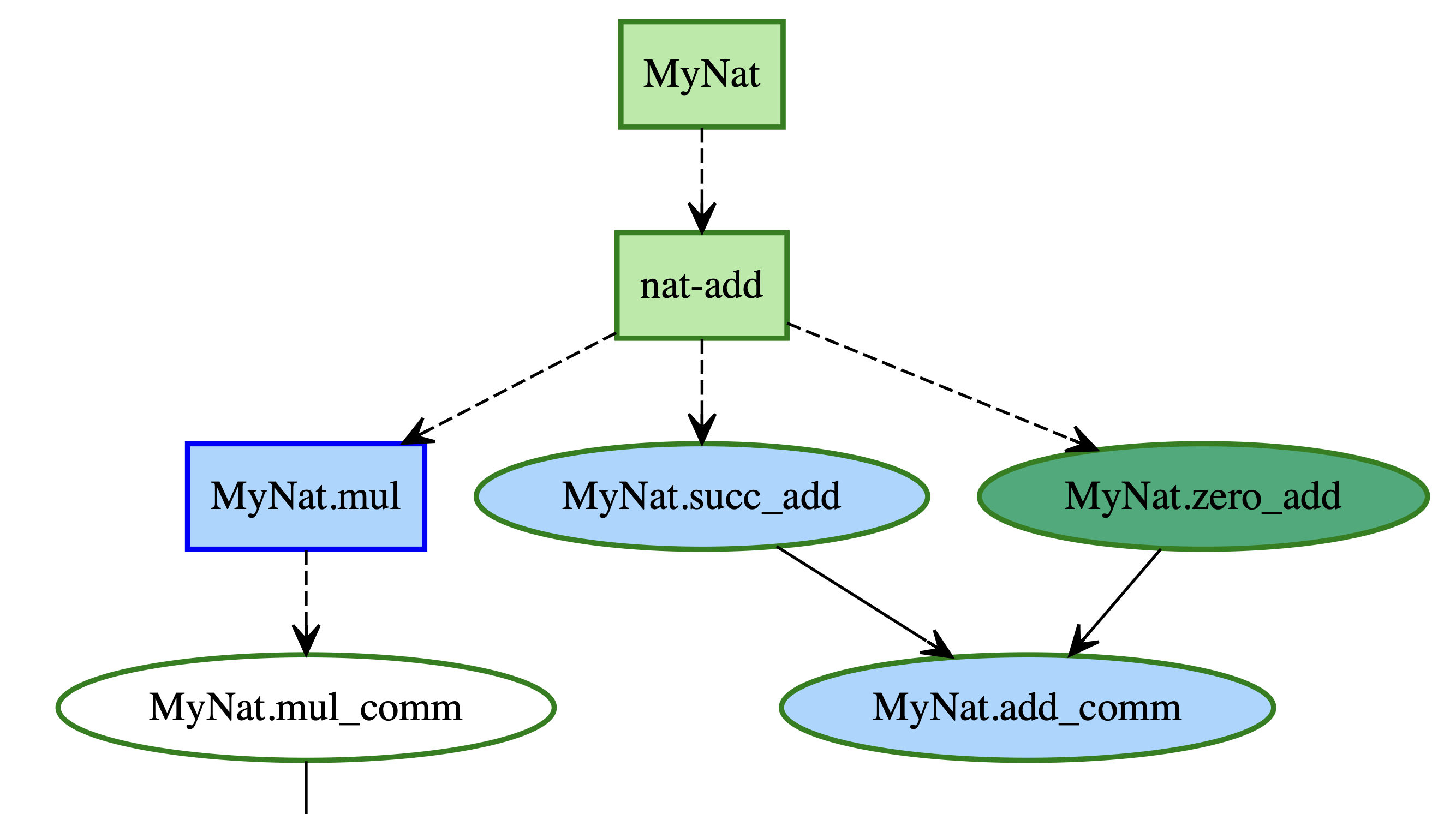}
      \caption{Dependency graph.}
  \end{subfigure}
  \caption{The blueprint generated from the example Lean file (\Cref{lean-example}).}
  \label{fig:example-blueprint}
\end{figure}

The dependency graph of the generated blueprint\footnote{See the \href{https://hanwenzhu.github.io/LeanArchitect-example/blueprint}{hosted blueprint}.} is visualized in~\Cref{fig:example-blueprint}. Note the automatically inserted statements and proofs, and the relations between the nodes.

\subsection{Implementation}

We briefly introduce the implementation of \leanarchitect. However, note that since writing this section, the \leanarchitect implementation has changed and will continue to change for new features, performance, and bug fixes.

At its foundation, \leanarchitect is built upon the following \leancode{Node} datatype, which represents a \leancode{@[blueprint]}-tagged Lean declaration with relevant metadata.

\begin{lstlisting}
/-- The statement or proof of a node. -/
structure §\declname{NodePart}§ where
  /-- The natural language description of this part. -/
  text : String
  /-- The specified set of nodes that this node depends on, in addition to inferred ones. -/
  uses : Array Name
  /-- The set of nodes to exclude from `uses`. -/
  excludes : Array Name
  /-- Additional LaTeX labels of nodes that this node depends on. -/
  usesLabels : Array String
  /-- The set of labels to exclude from `usesLabels`. -/
  excludesLabels : Array String
  /-- The LaTeX environment to use for this part. -/
  latexEnv : String

/-- A theorem or definition in the blueprint graph. -/
structure §\declname{Node}§ where
  /-- The Lean name of the tagged constant. -/
  name : Name
  /-- The LaTeX label of the node. Multiple nodes can have the same label. -/
  latexLabel : String
  /-- The statement of this node. -/
  statement : NodePart
  /-- The proof of this node. -/
  proof : Option NodePart
  /-- The surrounding environment is not ready to be formalized, typically because it requires more blueprint work. -/
  notReady : Bool
  /-- A GitHub issue number where the surrounding definition or statement is discussed. -/
  discussion : Option Nat
  /-- The short title of the node in LaTeX. -/
  title : Option String
\end{lstlisting}

Nodes are stored in a persistent environment extension, mapping the name of each \leancode{@[blueprint]}-tagged constant to the underlying node.

\begin{lstlisting}
/-- Environment extension that stores the nodes of the blueprint. -/
initialize §\declname{blueprintExt}§ : NameMapExtension Node ←
  registerNameMapExtension Node
\end{lstlisting}

A less common but useful feature is for a blueprint ``node'' to correspond to different Lean declarations \leancode{a} and \leancode{b}. In the output, the node would have \latexcode|\lean{a, b}|. To enable this feature, the user would specify the same \LaTeX{} label for different Lean declarations (which still technically correspond to different Lean \leancode{Node}s). We maintain a mapping from a \LaTeX{} label to the set of nodes with such a label, as an environment extension alongside \leancode{blueprintExt}. During the output, the nodes with the same label are merged to a single \LaTeX{} environment. This allows uses such as interaction with Mathlib's \leancode{to_additive}:

\begin{lstlisting}
@[to_additive (attr := blueprint "my-label")]
theorem §\declname{mul\_theorem}§ := sorry
\end{lstlisting}

which defines a blueprint node with Lean declarations \leancode{mul_theorem} and \leancode{add_theorem}.

We then define an attribute \leancode{@[blueprint]}. Upon tagging a Lean declaration, the attribute constructs a new \leancode{Node} with fields populated by the configuration given to \leancode{@[blueprint]}. The \leancode{@[blueprint]} attribute specifically has the following configuration options:

\begin{lstlisting}
@[blueprint
  "latex-label"             -- The LaTeX label to use for the node (default: Lean name)
  (statement := /-- ... -/) -- The statement of the node in LaTeX
  (hasProof := true)        -- If the node has a proof part (default: true if the node is a theorem)
  (proof := /-- ... -/)     -- The proof of the node in LaTeX (default: the docstrings in proof tactics)
  (uses := [a, "b"])        -- The dependencies of the node, as Lean constants or LaTeX labels (default: inferred)
  (proofUses := [a, "b"])   -- The dependencies of the proof of the node, as Lean constants or LaTeX labels (default: inferred)
  (title := /-- Title -/)   -- The title of the node in LaTeX
  (notReady := true)        -- Whether the node is not ready
  (discussion := 123)       -- The discussion issue number of the node
  (latexEnv := "lemma")     -- The LaTeX environment to use for the node (default: "theorem" or "definition")
]
\end{lstlisting}

Then we add the node with the above data to \leancode{blueprintExt}.

We also define two auxiliary tactics for writing proofs that more elegantly integrate with \leanarchitect. The first is allowing docstrings to prepend a tactic, such as \leancode{/-- Proof by induction on $b$. -/ induction b with ...}. Such docstrings will be concatenated and become the ``proof text'' of the node. The second is \leancode{sorry_using [a, b]}, which acts like \leancode{sorry} as a proof placeholder, but also allows for specifying constants used by the would-be proof (achieving the same effect as \leancode{proofUses}).

The \emph{blueprint content} of a module is the content of a Lean module to be converted to \LaTeX{}, and it is defined as the ordered list of \leancode{Node}s in the module. Users may also use \leancode{blueprint_comment /-- ... -/} to manually write raw \LaTeX{} to the blueprint content. Suppose Lean declaration \leancode{MyNat.add_comm} is tagged with \leancode{@[blueprint "thm:add-comm"]}. We specify the rules for converting this node into \LaTeX{} as follows.

\begin{itemize}
\item \latexcode|\label| is \latexcode{thm:add-comm}
\item \latexcode|\lean| is \leancode{MyNat.add_comm}
\item \latexcode|\uses| is the automatically inferred dependencies of the declaration, plus or minus any manually specified dependencies or exclusions. Automatic collection of used constants uses the same logic as \leancode{#print axioms} in Lean core, but it collects either dependencies tagged with \leancode{@[blueprint]} or axioms. The \latexcode|\uses| of the statement are the dependencies of the declaration type (and if a definition, its value), while the \latexcode|\uses| of the proof (if a theorem) are the dependencies of the declaration value.
\item \latexcode|\leanok| of the statement (resp.\ proof) is added if \leancode{sorryAx} is not in the inferred Lean dependencies above. For example, if the proof of \leancode{MyNat.add_comm} depends on \leancode{sorry}, \latexcode|\leanok| is not added to the proof; but if its proof \emph{only} depends on \leancode{MyNat.succ_add} which in turn depends on \leancode{sorry}, then \latexcode|\leanok| \emph{is} added to denote a completed proof formalization pending on an earlier unformalized lemma.
\item \latexcode|\mathlibok| is added to the statement (signifying the declaration is already in Mathlib) if \leancode{MyNat.add_comm} is defined in a module under \leancode{Init}, \leancode{Std}, \leancode{Batteries}, or \leancode{Mathlib}. To tag such a theorem, the user can write \leancode{attribute [blueprint] MyNat.add_comm}. This automatically differentiates if a theorem is already in Mathlib, easing management work for humans upstreaming results.
\item The text of the statement is given by \leancode{(statement := /-- ... -/)} in the attribute, and the proof by \leancode{(proof := /-- ... -/)} or the proof docstrings.
\item The statement title (\latexcode|\begin{theorem}[title] ... \end{theorem}|) is given by \leancode{(title := /-- ... -/)}.
\item Similarly for other metadata such as \latexcode{\discussion} and \latexcode{\notready}, which are utility features in \leanblueprint.
\end{itemize}

The output \LaTeX{} defines two macros:
\begin{itemize}
\item \bplatexcode|\inputleannode{thm:add-comm}|, which expands to a node in the \LaTeX{} blueprint with the above information, as shown in~\cref{sec:latex-export};
\item \bplatexcode|\inputleanmodule{MyModule}|, a higher-level command which expands to the \LaTeX{} blueprint content of the entire module \latexcode{MyModule}.
\end{itemize}
These macro definitions are saved to separate files in the \LaTeX{} build artifacts directory \bashcode{.lake/build/blueprint}. In the \LaTeX{} blueprint, the user may import the macros and write \bplatexcode|\inputleannode{...}| to enter a theorem in the blueprint.

Parallel to doc-gen4, \leanarchitect provides an internal \leancode{lean_exe} executable \leancode{extract_blueprint} that outputs the \LaTeX{} artifacts of a single module to a file in \bashcode{.lake/build/blueprint}. \leanarchitect then specifies a module facet\footnote{See the \href{https://lean-lang.org/doc/reference/latest/Build-Tools-and-Distribution/Lake}{Lake documentation}.} \leancode{blueprint} (\bashcode{lake build My.Module:blueprint}), that calls this executable and registers the output file, allowing Lake’s incremental build mechanism to only build the file if it is not up-to-date. We define a library facet \leancode{blueprint} (\bashcode{lake build my-library:blueprint}) that outputs the main header file, and a package facet \leancode{blueprint} (\bashcode{lake build :blueprint}) that runs the library facet on all libraries.

The \leancode{blueprintConvert} script primarily uses Python regular expression parsing to read the nodes in the existing blueprint, and calls Lean to obtain the locations of each node in Lean. Then it inserts \leancode{@[blueprint]} tags at appropriate locations in the code (while respecting existing attributes like \leancode{to_additive}). For a Lean declaration imported from another project (usually, from Mathlib), the script inserts \leancode{attribute [blueprint] mathlib_node} at an appropriate location: immediately before the first node (in topological order) that depends on this node, or the root file if there is no such location. Finally, the script replaces the existing \latexcode|\begin{theorem}\end{theorem}| nodes in \LaTeX{} with \bplatexcode|\inputleannode{}|. The conversion script is invoked by a Lake script called by \bashcode|lake script run blueprintConvert|. Among the customizable options are: whether to convert only the nodes with \latexcode|\lean| (default) or all nodes, whether to remove the \latexcode|\uses| information for \latexcode|\leanok| nodes and let \leanarchitect infer it (default) or not, and formatting options for docstrings.

\end{document}